\documentclass{aa}
\usepackage{natbib,psfig}
\begin{document}

\title{The extreme ultraviolet excess emission in five clusters of galaxies
revisited}
\titlerunning{Excess EUV emission in five clusters}

\author{Florence Durret\inst{1} \and 
Eric Slezak \inst{2} \and
Richard Lieu\inst{3} \and 
Sergio Dos Santos\inst{1,4} \and 
Massimiliano Bonamente\inst{3}}
\institute{Institut d'Astrophysique de Paris, 98bis Bd Arago, F-75014 Paris, France \and
Observatoire de la C\^ote d'Azur, B.P. 4229, F-06304 Nice  Cedex 4, France
\and
Department of Physics, University of Alabama, Huntsville AL 35899, USA
\and
Department of Physics and Astronomy, University of Leicester, University Road, 
Leicester LE1 7RH, UK}

\offprints{F. Durret, durret@iap.fr}

\date{Accepted 15-05-02, Received 1-05-02, In original form 18-02-02}

\abstract{ Evidence for excess extreme ultraviolet (EUV) emission over
a tail of X-ray gas bremsstrahlung emission has been building up
recently, but in some cases remains controversial, mostly due to the
moderate quality of the EUV data. In order to improve the signal to
noise ratio in the EUV, we have performed the wavelet analysis and
image reconstructions for five clusters of galaxies observed both at
EUV and X-ray energies with the EUVE and ROSAT satellites
respectively. The profiles of the EUV and X-ray reconstructed images
all differ at a very large confidence level and an EUV excess over a
thermal bremsstrahlung tail is detected in all five clusters (Abell
1795, Abell 2199, Abell 4059, Coma and Virgo) up to large radii. These
results, coupled with recent XMM-Newton observations, suggest that the
EUV excess is probably non thermal in origin.  } \maketitle

\section{Introduction}

The Extreme Ultraviolet Explorer (EUVE) has detected emission from a
few clusters of galaxies in the $\sim$70-200 eV energy range. By order
of discovery, these were: Virgo \citep{Lieu2,Berghofer1}, Coma
\citep{Lieu1}, Abell 1795 \citep{Mittaz}, Abell 2199 \citep{Bowyer},
Abell 4059 \citep{Berghofer2} and Fornax (Bowyer et al. 2001). The
problem of the physical origin of the EUV emission was immediately
raised. Excess EUV emission relative to the extrapolation of the X-ray
emission to the EUVE energy range was detected in several clusters,
suggesting that thermal bremsstrahlung from the hot ($\sim 10^8$K) gas
could not be entirely responsible for the EUV emission. Although there
has been some controversy on the reality of this EUV emission, it now
seems well established for the following clusters: Coma \citep{Lieu1},
Virgo \citep{Lieu2,Berghofer1,Bonamente1}, Abell 1795
\citep{Mittaz,Bonamente1}, Abell 2199 \citep{Bowyer,Kaastra,Lieu5} and
Abell 4038 \citep{Bowyer}.

This excess EUV emission can be interpreted as due to two different
mechanisms: thermal radiation from a warm ($10^5-10^6$K) gas, as first
suggested by Lieu et al. (1996a), or inverse Compton emission of
relativistic electrons either on the cosmic microwave background or on
stellar light originating in galaxies (Hwang 1997; Bowyer \&
Bergh\"ofer 1998; Ensslin \& Biermann 1998; Sarazin \& Lieu 1998;
Ensslin et al. 1999), or both.  Note that although the existence of a
multiphase intracluster medium is plausible \citep{Lieu4,Bonamente2},
there are several difficulties with the thermal model. The most
serious one is that since gas in the temperature range $10^5-10^6$K
cools very rapidly, a source of heating for this gas is necessary, and
only one model has been proposed to account for this gas heating
(Fabian 1997). Besides, the non-detection by the Far Ultraviolet
Spectroscopic Explorer of the OVI $\lambda\lambda$1032,1038 lines in
the spectra of the Coma and Virgo clusters could exclude the presence
of warm gas at a 2$\sigma$ level \citep{VDD}. 
On the other hand, non thermal models \citep{Sarazin,
Bowyer1,Lieu3,Ensslin,Atoyan,Brunetti} seem able to account for the
excess EUV emission, but require very high cosmic ray pressure.

Following a controversy that the extended EUV emission from some
clusters could be simply an artefact of bad background subtraction
(Bowyer et al. 1999; Lieu et al. 1999b) we have performed a wavelet
analysis and reconstruction of the EUVE images of several
clusters. Our aims were to confirm the detection of EUV emission as
far as possible from the cluster center, to derive accurate EUV and
X-ray profiles and to confirm the existence and radial distributions
of the soft excesses over thermal bremsstrahlung in these clusters.

We present here the wavelet reconstruction of the images of these five
clusters, both in the EUV (EUVE images) and in X-rays (ROSAT PSPC
images).  Abell 4038 was also analyzed but was discarded due to its
low signal to noise ratio in the EUV. We then compare the EUV and
X-ray profiles of the images thus obtained, arguing that if both
emissions have the same physical origin it should not be possible to
distinguish the shapes of these profiles. In fact, we will see that
they are found to be different in all clusters, and that soft excesses
over thermal bremsstrahlung are detected in all five clusters with
various radial distributions.

The data and method are presented in Sect.~2. Results are presented
in Sect.~3 and discussed in Sect.~4.

\section{The data and image analysis}

\subsection{The data}

Five clusters were observed in the EUV range with the EUVE satellite
and in X-rays with the ROSAT PSPC satellite (archive data).  The
exposure times and main cluster characteristics are given in
Table~\ref{tbl:mainchar}.

\begin{table*}
\centering
\caption{Exposure times and main cluster characteristics.}
\begin{tabular}{lrrclrrc}
\hline
Cluster & EUVE total  & ROSAT  exp. & ROSAT & Redshift & kT$_{\rm X}$ & Scaling & Cooling\\
         & exp. time (s)  &  time (s)   &  ObsId       &          & (keV)        & (kpc/superpx) & flow\\
 \hline		    
 A 1795  & 158689    &   33921     & rp800105     & 0.063    & 5.9          & 32.7    & yes \\
 A 2199  &  93721    &   34633     & rp800644     & 0.0299   & 4.1          & 15.8    & yes \\
 A 4059  &  145389   &    5225     & rp800175     & 0.0478   & 4.0          & 25.0    & yes\\
 Coma    &   60822   &   20112     & rp800005     & 0.023    & 8.7          & 12.2    & no \\
 Virgo   &  146204   &    9135     & rp800187     & 0.003    & 2.4          &  1.6    & weak \\
\hline
\end{tabular}
\label{tbl:mainchar}
\end{table*}

In order to have comparable spatial resolutions in the EUV and X-rays,
the EUV images were rebinned 4$\times$4, leading to a ``superpixel''
size of 0.3077 arcminutes (18.5 arcseconds). X-ray images were also
rebinned to the same 18.5 arcsec pixel size in order to allow a direct
comparison of the EUV and X-ray profiles.

The linear scale per superpixel at the cluster distance, estimated
with H$_0=50$~km~s$^{-1}$~Mpc$^{-1}$ and q$_0$=0 is given in Col. 7
of Table~\ref{tbl:mainchar}.

\subsection{The EUVE data}

The EUVE/Deep Survey photometer (DS) images were extracted from the
raw event lists with the customary linear bin size of 13 pix/arcmin.
The rectangular shape of the Lex/B filter ($\sim$70-200 eV) results in
images where the length of one side of the image greatly exceeds the
other side (see Fig.~1). When more than one observation was available
for each target, images were coadded to improve the S/N; the
rectangular shape often results in only partial overlap between the
exposures leading to diamond-shaped geometries (see Fig.~2). Without
taking into account any edge corrections, this peculiar shape of the
EUVE images may be a problem for any analysis involving an isotropic
distribution of the data at scales larger than about one-third the
size of the shorter side.

\begin{figure}
\centering
%\mbox{\psfig{figure=durret.fig1.ps,width=8.5cm}}
\caption[]{Virgo long exposure -- arrangement of boxes in
order to obtain the image ready for the wavelet reconstruction. Note that
the different boxes here are only illustrative of the algorithm we used. }
\label{fig:virgo}
\end{figure}

Providing that the image does not truncate the cluster emission in any
direction, this shape problem can be treated as a classical edge
effect. A simple solution is then to use the asymptotic background
values to obtain a simulated larger square image. From a practical
point of view only a small truncation of the data occurs in our images
along the shorter axis. Its impact on an azimuthal profile can
therefore be considered as negligible and we thus decided to apply the
above procedure to modify the geometry of the EUVE images. We now give
details on the required background estimation.

Fig.~\ref{fig:virgo} shows the rebinned image of a long observation of
the Virgo cluster. The cluster is clearly visible in the central area
of the field of view. Three rectangles are superimposed on the raw
image. The largest one, denoted by (A,B,C,D) follows the outer
detector shape closely. Its edges are detached from the detector
boundaries to avoid the steepening of the background near the edge,
which would bias the noise estimation \citep[see the figure 2
of][]{Lieu6}.  In this box, we define two rectangles (denoted $r_1$
and $r_2$), one towards each edge, as far away from the central source
as possible. Both $r_1$ and $r_2$ are assumed to contain only noise
and to be representative of the noise level far from any celestial
source. This is ensured because the centers of both rectangles lie at
$\sim 40$~arcmin of the central source, a distance notably greater
than the detection radius of the extended EUVE cluster emission, as
seen with classical methods. The second order moment of the count
distribution is used to estimate the level of the noise in each of
these rectangles. The noise level in the large box is then computed as
the mean of the noise levels in both rectangles. Once this has been
done, each pixel outside the larger box is replaced by a random number
drawn from a Poisson distribution, the variance of which is obtained
by adding in quadrature the variances in each of the rectangles. The
accuracy of the noise estimation is readily observable in the product
image: if the noise is underestimated, the central box will obviously
rise against the rest of the image, with its boundaries showing a
clear discontinuity. On the other hand, if the noise has been
overestimated, the outside parts of the new image will rise against
the central box, again showing a discontinuity on the boundaries.

\begin{figure}
\centering
%\mbox{\psfig{figure=durret.fig2.ps,width=8.5cm}}
\caption[A1795 polygons]{Abell 1795 arrangement of boxes in order to obtain
the image prior the wavelet reconstruction in the case of a co-added
image. The central circle has a radius of 16 arcmin, which we will take as
the limit of our analysis.
 } 
\label{fig:a1795box}
\end{figure}

\begin{figure}
\centering
%\mbox{\psfig{figure=durret.fig3.ps,width=7.5cm}}
\caption[A1795 polygons]{Enlarged and rebinned image of Abell 1795.
The total size of the image is 140$\times$140 arcmin$^2$.}
\label{fig:a1795addnoise}
\end{figure}

When more than one EUVE observation of a cluster was available, we
used the co-added images in order to increase the signal to noise
ratio. Fig.~\ref{fig:a1795box} illustrates the case of Abell 1795.
This image is typical of the co-added images we dealt with in the case
of other clusters. The difference in roll-angle of the satellite in
both observations is apparent. The central intersection of both
observations obviously has the highest signal to noise ratio. It
corresponds to a region indicated by the circle of radius 16 arcmin in
Fig. \ref{fig:a1795addnoise}, which we will take as the radial limit
for our analysis.  For each of the separate observations, we drew
again a box closely following the outer detector shape. Both boxes are
labelled (A,B,C,D) and (A$^\prime$,B$^\prime$,C$^\prime$,D$^\prime$)
in Fig.~\ref{fig:a1795box}. In each of the boxes, we define two small
rectangles, where the noise is evaluated. The noise level in a box is
computed as the mean of the noise levels in both rectangles. Once this
has been done for each box, each pixel outside the intersection of the
boxes (the central diamond) is again replaced by a random number drawn
from a Poisson distribution, as discussed before. Once again, the
under or overestimation of the noise will be clearly visible. An
example of such an image is shown in Fig.~\ref{fig:a1795addnoise}.

\subsection{The ROSAT data}
\label{sec:rosatdata}

The {\sf ROSAT\/} X-ray data for each cluster were obtained from the
HEASARC web archive\footnote{http://heasarc.gsfc.nasa.gov/}. Owing to
the higher sensitivity, we preferred PSPC to HRI data and pointed
observations to all-sky survey scanning mode observations.  When
multiple PSPC observations were available, we always used the one with
the smallest offset radius compared to the center of the cluster, and
then the longest one.

S. Snowden's Extended Source Analysis Software
(ESAS\footnote{ftp://legacy.gsfc.nasa.gov/rosat/software/fortran/sxrb})
was used to perform the data reduction, {\it i.e.\/,} rejection of
high-background times, modelling of the different background types,
energy-dependent background subtraction, exposure and vignetting
corrections (see Snowden et al. 1994).  We adopted a conservative
value of $170 \, {\rm cts~s^{-1}}$ for the maximum Average Master Veto
rate allowed.  We then carefully examined the light curves of the
total counts in the entire image per energy band, and checked that no
short time scale glitches were present.  We limited our analysis to
the [0.5,2.0] keV energy band (bands R4-R7 as defined by
Snowden). This preliminary reduction produced a surface brightness
image with $512 \times 512$ pixels, 15 arcsec per pixel (roughly the
FWHM of the {\sf PSPC} PSF at 1~keV), as well as an exposure
image for each of the bands and a background image. The image was then
divided by the mean exposure map. We thus obtained a $512 \times 512$
pixel image, in units of counts s$^{-1}$ pixel$^{-2}$, which was
then rebinned to 18.5$\times$18.5 arcsec$^2$ pixels to match the EUV
superpixel size.

The ROSAT images do not present the same problems as the EUVE ones,
because in this case, the geometry is circular around the central
cluster (all our observations were pointed). The only geometrical
constraint arises from the presence at an offset angle of
approximately 20 arcmin of the structure support of the
telescope. However, since this value is smaller than the EUVE limiting
radius, it does not place any further constraint here. We therefore
just extracted a square image inscribed in the 20 arcmin radius circle.

\subsection{Wavelet analysis and reconstruction}

The comparison between the EUVE observed surface brightness profiles
of clusters and their ROSAT X-ray predicted counterparts is the key
step in our analysis. However a direct comparison of the X-ray and
EUVE data would provide misleading results due to the noise in the raw
data. To remove most of this noise while preserving information on the
astronomical emission at all spatial scales of interest we decided to
apply part of the wavelet vision model described in detail in Ru\'e
\& Bijaoui (1997).

Let us hereafter briefly summarize the main steps of the image
processing involved. First, a discrete wavelet transform of the image
$I_0$ is performed using the so-called {\sl \`a trous} implementation,
which is nothing more than the classical multiresolution algorithm of
Mallat (1989) but without any decimations since ``holes'' are introduced
in the required convolutions. The initial sampling is kept for each
scale; this insures shift-invariance and allows straightforward
interscale computations. The scaling function leading through a
dilation factor $a$ to the involved sets of low-pass and band-pass
(wavelets) filters is a cubic B-spline (defined in the Fourier space
as {\rm sinc}$^4(\omega)$), very well suited for enlightening gaussian-like
patterns of various sizes in an image. Owing to the properties of the
underlying convolutions a stationary white noise signal leads to a
statistical distribution of the wavelet coefficients with a zero mean
value at each computed scale $a_j$ and a standard deviation decreasing
as the spatial scale increases. Structures in the transform which are
related to objects in direct space can then be identified and located
by selecting high enough wavelet coefficients at each scale.

Removing noise from the data without any loss in resolution can thus
be achieved quite easily in wavelet space by keeping only those
coefficient values with a probability to be due to a chance
fluctuation of the noise process lower than a chosen decision
level. This second step of the noise removal process requires the
knowledge of the probability density function of the coefficients at
scale $a_1=1$~pixel (here 18.5 arcsec) for the noise process and of the
decreasing law of the rms scatter $\sigma$, the asymptotic value of
which appears to be 0.5 for a bidimensional image $I_0$ and a dyadic
scale analysis ($a_j=2^{j-1}, j\in {\cal Z}$)~: $\sigma_j= \sigma_1 /
2^{j-1}$. In case of Gaussian noise, the probability density function
is easily obtained from a histogram of the wavelet coefficients
providing that this first scale of one pixel in size is noise
dominated. Due to the stationarity of the Gaussian process, the
selection of the so-called statistically significant coefficients
$\nu_j$ at scale $a_j$ can then be done using the same k-$\sigma_j$
threshold. We end up in this way with a subset $\nu^{(0)}$ of
coefficients at different scales, the pixels of which define what is
called hereafter the multiscale significant mask $M_0$. In case of
Poisson noise stationarity can be obtained by applying an Anscombe
(1948) transform on the data~: $f(x)=2\sqrt{x+ 0.375}$. When the mean
parameter $\mu$ of the Poisson process is greater than about 5, this
transform indeed leads to a new process with a very small bias in the
expectation value (${ <f(x)>-f(\mu) \over f(\mu)} <2\%$) and a nearly
perfect variance stabilization. The same decision scheme as in the
Gaussian case can therefore be used for defining the multiscale
significant mask from the wavelet transform of this image with a
constant variance. Only second order errors in the wavelet
coefficients would arise at the very first scales in areas with very
low counts. This is not a matter of concern for the present
application since the computation of a correct azimuthal brightness
profile necessitates only a proper account of the medium and
large-scale features in these faint regions.

The third and final step is to restore a positive image $I$ where
the noise is strongly reduced from the subset $\nu^{(0)}$ of the
significant wavelet coefficients of $I^{(0)}\equiv I_0$. This is an inverse
problem which has in general an infinity of solutions and the choice
among these solutions must be done with respect to a regularization
constraint. To do so, an iterative conjugate gradient algorithm is used.
This algorithm minimizes the distance $\|\nu^{(0)} -\nu^{(n)} \|$ where
$\nu^{(n)}$ is the restriction to $M_0$ of the wavelet transform of
the solution $I^{(n)}$ at step $n$. Stating A the operator associated
to the wavelet transform followed by the projection onto $M_0$ and
\~A the joint operator, $I^{(n+1)}=$~\~A$(\nu^{(n)})$ is obtained from
the previous solution by~: $I^{(n+1)}= I^{(n)} + \alpha^{(n)} I_r^{(n)}$
where the residual image $I_r^{(n)}$ is defined as
\~A$(\nu_r^{(n)}) + \beta^{(n)} I_r^{(n-1)}$ 
with $\nu_r^{(n)}= \nu^{(0)} - $~A$(I^{(n)})$~;
$\alpha$ and $\beta$ are convergence parameters.
Convergence ($\frac{\|\nu^{(0)} -\nu^{(n)} \|}{\|\nu^{(0)} \|} < 1\%$) 
is reached after very few steps, leading to an image $I$ without any
discontinuities despite the hard thresholding performed in the wavelet
space. Note also that any flat instrumental or astronomical component
or gradient in the initial image is automatically removed in image $I$
taking benefit of the mathematical properties of the wavelet function
related to our scaling function.

We applied such an image processing to both our EUVE and ROSAT square
images considering seven or eight spatial scales for the wavelet
transform according to the size of the image. Pure Poisson noise was
considered in both cases. This is not strictly the case for the ROSAT
data because of the preprocessing of the raw counts (exposure
correction, modelled background subtraction), but this approximation
appears to be good enough for detecting correctly all the large-scale
components we are interested in. Significant wavelet coefficients were
thus selected according to a probability of false detection of
4$\sigma$, 3.5$\sigma$ and 3$\sigma$ for scales 1 (18.5 arcsec), 2
(37 arcsec) and 3 (74 arcsec) and above respectively. The rather high
levels for the first two scales automatically eliminate most of the
small-scale structures if any (these might distort the overall shape
of the brightness profile or might have arisen from an error in the
probability density function of the noise or from a loss in control
over the bias in the Anscombe transform in regions with very few counts).
We have checked in several cases that the profiles do not change
significantly when the thresholding for scales 1, 2 and 3 is modified.

\subsection{Radial profiles}

After wavelet reconstructing both EUVE and ROSAT images for each
cluster, we proceed to compare the decrease of the surface brightness
in the two observational bands.  Profiles were thus derived from the
2D reconstructed images within concentric elliptical rings using the
STSDAS.ANALYSIS. ISOPHOTE.ELLIPSE task in IRAF. The cluster
ellipticities and major axis position angles were estimated from the
X-ray images, which have better signal to noise ratio, except for Coma
because it fills the 20 arcmin PSPC ring and its ellipticity had to be
derived from the EUVE reconstructed image. The ellipticities and major
axis position angles were then fixed to be the same in the EUV and
X-rays, in order to measure intensities in the same spatial regions.

In order to see how the instrument point spread functions (PSF)
could influence our results, we also analyzed a point source in a
similar way: the star $\gamma$Tau, for which both EUV and X-ray data
were available. The EUV and X-ray exposure times for this star were
89924s and 8547s respectively, and the ROSAT PSPC image of $\gamma$Tau
was divided by its corresponding exposure map, as for clusters. The
EUV and X-ray images were rebinned to a pizel size of 18.5$\times$18.5
arcsec$^2$ as previously, and the EUV and X-ray profiles were then
derived in a similar way. Note that this point source is about 15
arcmin offset from the detector center in the ROSAT image, at a radius
where the PSF of PSPC is already degraded. Therefore, the PSF effect
in X-rays will in fact always be smaller or equal than what we measure
for $\gamma$ Tau. This is not the case for EUVE, where the point
source is close to the detector center.

The problem of a possible non-flatness of the background was already
addressed in the case of EUVE by Lieu et al. (1999c). A comparable
wavelet analysis was performed on a blank field totalling 85 ksec
exposure time, and the backgroud was found to vary by at most 10\%
within the central circle of 16 arcmin radius defined above taken into
account here (see Fig. 1 by Lieu et al. 1999c). The mean value of
another blank field observed for 21 ksec is found to be 1.21
counts/s/pixel$^2$, excluding the outermost regions (that is in a
rectangle similar to ABCD from Fig. \ref{fig:virgo}), which is comparable
to the mean background value in Abell 1795 (1.19 counts/s/pixel$^2$).

In the case of the X-ray images, they were divided by their
corresponding exposure maps, and therefore the non-flatness of the
ROSAT PSPC detector should be corrected for. We are therefore
confident that the non-flatness of the detector backgrounds cannot
change influence our results by more than a few \%.

Error bars on all profiles are those directly given by the ELLIPSE
task. However, since we were concerned that they might be
underestimated, we also computed them from the Poisson fluctuation in
the number of counts in each ring ; these error bars were found to be
commensurate with those given by ELLIPSE (the rms fluctuation of the
pixel intensity within each concentric ellipse). For reconstructed
images, they are typically smaller than $\pm 0.01$ in logarithmic
scale, i.e.  too small to be clearly visible on
Figs. \ref{fig:a1795profil} to \ref{fig:virgoprofil}).

\section{Results}

\subsection{EUV and X-ray profiles}

For co-added images, we do not expect to detect any structure outside
a certain radius since the outer part of the image is purely
noise. The central circle in Fig.~\ref{fig:a1795box} shows a
conservative estimate of this radius, i.e. the half width of the EUVE
detector $\sim$16 arcmin, which we will take as the limit of our
analysis.

\begin{figure}[h]
\centering
\mbox{\psfig{figure=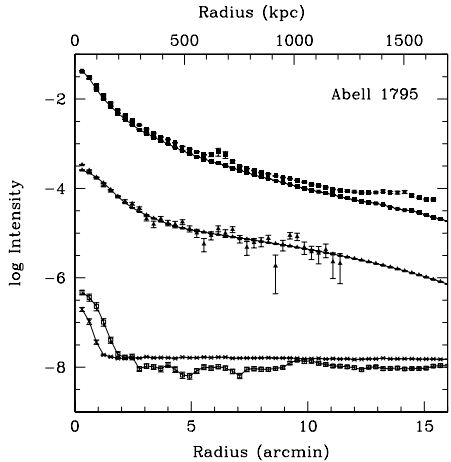,width=6.5cm}}
\caption[]{EUV and X-ray profiles of Abell 1795, obtained directly
from the raw data (points) and after wavelet analysis and
reconstruction (points with connecting lines). As in the four
following figures, the EUVE data are drawn with triangles (bottom
curves) and the X-ray data with filled squares (top curves); the
radius is expressed in arcminutes (bottom) and kpc (top); intensities
are in counts/s/pixel$^2$ after background subtraction (1 pixel
is 18.5$\times$18.5 arcsec$^2$). The error bars are drawn, but for the
profiles derived from the wavelet reconstructed images they are too
small to be clearly visible (typically smaller than $\pm 0.01$ in
logarithmic scale for Figs. \ref{fig:a1795profil} to
\ref{fig:virgoprofil}). The EUV (crosses) and ROSAT PSPC (empty
squares) PSFs drawn from the images of the point source $\gamma$Tau,
shifted by an arbitrary value to make the figure clearer, are shown at
the bottom for comparison.  }
\label{fig:a1795profil}
\end{figure}

\begin{figure}[h]
\centering
\mbox{\psfig{figure=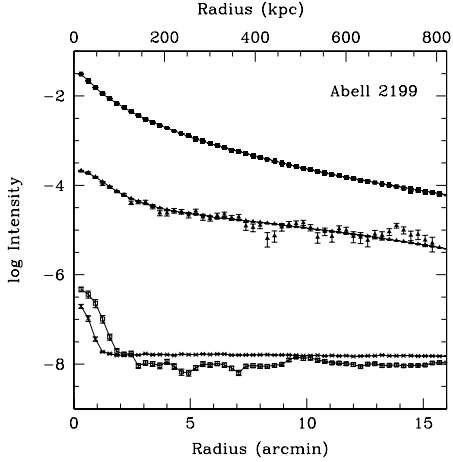,width=6.5cm}}
\caption[]{Same as Fig.~\ref{fig:a1795profil} for Abell 2199. }
\label{fig:a2199profil}
\end{figure}

\begin{figure}[h]
\centering
\mbox{\psfig{figure=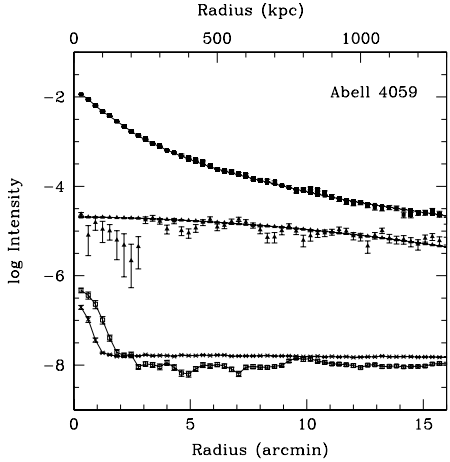,width=6.5cm}}
\caption[]{Same as Fig.~\ref{fig:a1795profil} for Abell 4059. }
\label{fig:a4059profil}
\end{figure}

\begin{figure}[h]
\centering
\mbox{\psfig{figure=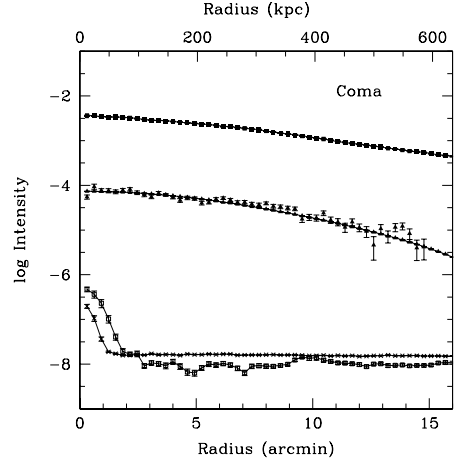,width=6.5cm}}
\caption[]{Same as Fig.~\ref{fig:a1795profil} for Coma. }
\label{fig:comaprofil}
\end{figure}

\begin{figure}[h]
\centering
\mbox{\psfig{figure=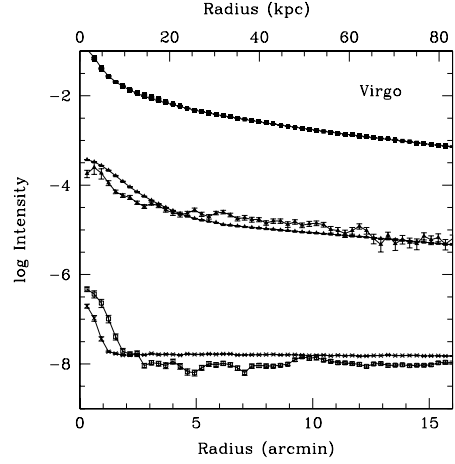,width=6.5cm}}
\caption[]{Same as Fig.~\ref{fig:a1795profil} for Virgo. }
\label{fig:virgoprofil}
\end{figure}

We first derived the profiles of the EUVE and X-ray images of all
clusters before and after wavelet analysis and reconstruction, as a
test to check that our method conserves the overall profile shape and
intensity level. The corresponding profiles are shown in
Fig.~\ref{fig:a1795profil} for Abell 1795. Note that all the
profiles drawn in Figs. \ref{fig:a1795profil} to \ref{fig:virgoprofil}
are obtained after subtracting the background. For the raw data, this
background is estimated as a mean value far from the edges of the
detector; for wavelet reconstructed images, the background is modelled
during the wavelet analysis itself and subtracted.
There is good consistency between the two EUVE curves on the one side
and between the two X-ray curves on the other side, implying that the
wavelet analysis and reconstruction applied here does not affect
fluxes, but simply provides a means of detecting cluster emission out
to a somewhat larger radius (as in the case of the EUVE data). The
small feature apparent in the X-ray profile at $\sim$6 arcmin is due
to a Seyfert galaxy superimposed on the cluster; it was removed from
the wavelet reconstructed image and therefore does not appear on the
corresponding profile. Note however that Abell 1795 is the only
cluster for which the raw and wavelet reconstructed X-ray profiles do
not coincide {\sl exactly} (in the other clusters they can hardly be
distinguished). Although we have tried various thresholding values for
the small scales and various background subtractions of the raw data
to see if they could coincide better, we have not been able to
superimpose them exactly.

The EUV and ROSAT PSPC PSFs drawn from the images of a point source
(see Sect. 2.5.) for comparison are also plotted in
Fig. \ref{fig:a1795profil}. They indicate that the PSFs have no
influence on the profile shapes for radii larger than about 1arcmin
and 2.5 arcmin in the EUV and X-rays respectively. Therefore any
features found in the profiles within regions smaller than these radii
are probably not real.

The EUV and X-ray radii of Abell 1795 are at least 16 arcmin,
corresponding to a physical extent of about 1700 kpc. This extent is
notably larger than the previously found value of 10 arcmin (Mittaz et
al. 1998, Bonamente et al. 2001b).

The EUV and X-ray radii of Abell 2199 extend to at least 16 arcmin, or
820 kpc (Fig.~\ref{fig:a2199profil}), comparable to the value of 15
arcmin measured by Kaastra et al. (1999) both with EUVE and Beppo-Sax.

In Abell 4059 (Fig.~\ref{fig:a4059profil}), the EUV and X-ray
emissions also reach 16 arcmin or 1300 kpc, much further than the 4-5
arcmin reported by Bergh\"ofer et al. (2000b).  The slopes of the
profiles in the EUV and X-rays differ notably, at least in the inner 8
or 10 arcmin. The difference in the central regions between the EUV
profiles derived from the raw and wavelet reconstructed images may be
due to the fact that structures only present at small scales were
eliminated, and therefore signal may be missing in the very center.
This cannot be due to the shape of the EUVE PSF, since it is flat
for radii larger than about 1 arcmin while this discrepancy is
observed up to a radius of about 3 arcmin. Such a behaviour is not
observed in Abell 1795, Abell 2199 or Coma, probably because they are
brighter.

The EUV and X-ray radii in Coma reach at least 16 arcmin
(Fig.~\ref{fig:comaprofil}), corresponding to 630 kpc and slightly
larger than reported by Bowyer et al. (1999).  The two profiles appear
quite similar at least up to a radius of about 10 arcmin. 

Virgo is the most nearby cluster in our sample. EUV and X-ray
emissions reach at least 16 arcmin (Fig.~\ref{fig:virgoprofil}), that
is slightly larger than detected by Lieu et al. (1996a) and
Bergh\"ofer et al. (2000a), but smaller than the 20 arcmin claimed by
Bonamente et al. (2001b); however, this extent corresponds to the very
small physical value of $\sim 80$ kpc, due to the proximity of
Virgo. Both profiles appear quite parallel except in the inner 4
arcmin; note however that since the X-ray PSF is not flat for
radii smaller than 2.5 arcmin, the difference between the EUV and
X-ray profiles in the inner 4 arcmin may not be real.

In order to quantify the probability for the EUV and X-ray profiles to
be similar, we performed a Kolmogorov-Smirnov (hereafter K-S) test on
the EUV and X-ray profiles drawn from the wavelet reconstructed
images, both in concentric elliptical annuli, and in ellipses with
increasing radius, after normalizing the EUV and X-ray profiles to
the same innermost pixel value.

In all clusters, the probability that the two profiles are issued from
the same parent population varies with radius: it is a few \% in the
very center (1.5 arcmin radius), and less than 2\% in concentric
annuli of width 1.5 arcmin (i.e. for radii between 1.5 and 3.0, 3.0
and 4.5 arcmin etc.).  If we consider ellipses of radii 3.0, 4.5, 6.0
arcmin etc. instead, the probability that the two profiles are issued
from the same parent population is smaller than 0.1\% in all
cases. Therefore even if the EUV and X-ray intensity profiles may
appear similar to the eye, as for example in Coma, it is clear that
statistically they are not.

\subsection{EUV to X-ray intensity ratios}

\begin{figure}[h]
\centering
\mbox{\psfig{figure=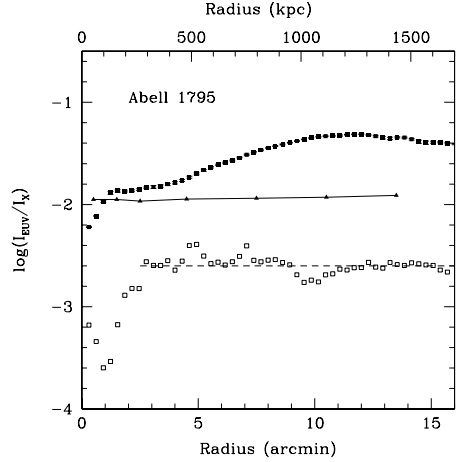,width=6.5cm}}
\caption[]{Ratio of the EUV to X-ray emission in Abell 1795 in
logarithmic scale. As in all following figures, squares show the
measured EUV to X-ray flux ratios (error bars are difficult to see on
the graphs since they are typically smaller than $\pm 0.01$ in
logarithmic scale, as for the other clusters in
Figs. \ref{fig:a2199rapp}-\ref{fig:virgorapp}).  Triangles indicate
the EUV to X-ray flux ratios in various concentric rings, expected
from the hypothesis that the EUV emission is the tail of the
bremsstrahlung emission accounting for the observed X-rays (see
text). The ratio of the EUV to X-ray PSFs is also drawn as empty
squares, shifted by an arbitrary value to make the figure clearer; to
guide the eye, the mean value of this ratio is indicated with a dashed
line.
}
\label{fig:a1795rapp}
\end{figure}

\begin{figure}
\centering
\mbox{\psfig{figure=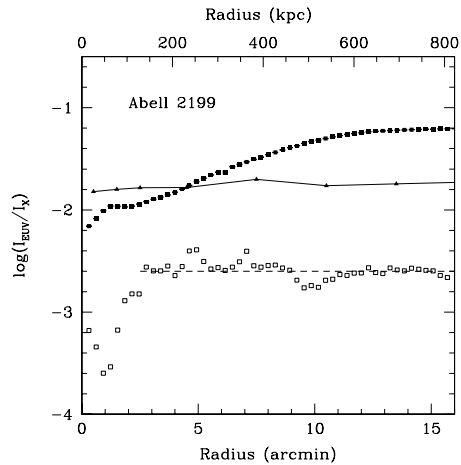,width=6.5cm}}
\caption[]{Same as Fig.~\ref{fig:a1795rapp} for Abell 2199. }
\label{fig:a2199rapp}
\end{figure}

\begin{figure}
\centering
\mbox{\psfig{figure=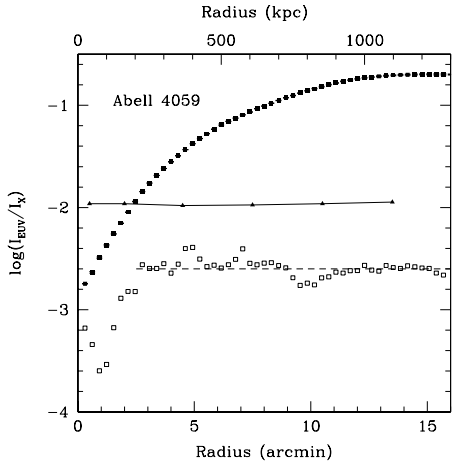,width=6.5cm}}
\caption[]{Same as Fig.~\ref{fig:a1795rapp} for Abell 4059. }
\label{fig:a4059rapp}
\end{figure}

\begin{figure}
\centering
\mbox{\psfig{figure=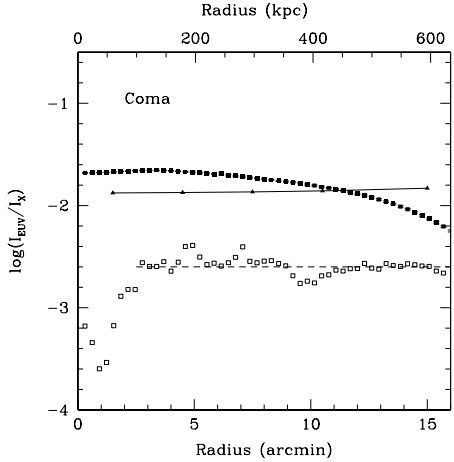,width=6.5cm}}
\caption[]{Same as Fig.~\ref{fig:a1795rapp} for Coma. }
\label{fig:comarapp}
\end{figure}

\begin{figure}
\centering \mbox{\psfig{figure=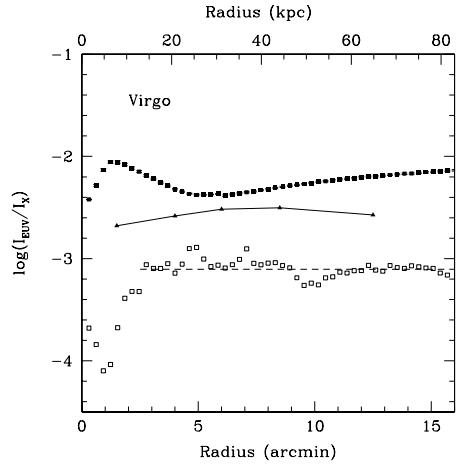,width=6.5cm}}
\caption[]{Same as Fig.~\ref{fig:a1795rapp} for Virgo. }
\label{fig:virgorapp}
\end{figure}

In order to compare the EUV to X-ray profiles, we computed the EUV to
X-ray intensity ratios in the same concentric ellipses; these are
shown in Figs. \ref{fig:a1795rapp} to \ref{fig:virgorapp}. The
ratio of the EUV to X-ray PSFs was also derived (see
Fig. \ref{fig:a1795rapp}). As expected from the individual EUV and
X-ray PSFs, this ratio is only constant for radii larger than 2.5
arcmin, while below it shows a strong dip.  Therefore the EUV to X-ray
ratios below this radius are not reliable. Such ratios have to be
compared to those expected in the hypothesis that the EUV emission is
the low energy tail of the hot ICM emission. We therefore employed
spatially resolved hot ICM models for each cluster, in order to
account for the effects of temperature and abundance gradients on the
predicted ratios. Spectra of all cluster regions were modelled with a
thin-plasma emission code (MEKAL in XSPEC, Mewe et al. 1985, 1986,
Kaastra 1992) modified by photoelectric absorption (WABS in XSPEC,
Morrison \& McCammon 1983). In detail, the hot ICM in Abell 1795 was
modelled with kT increasing from 3.5 keV at the center to 6 keV at
radii $\geq$ 3 arcmin, and abundances in the 0.3-0.5 solar range
(Briel \& Henry 1996); Abell 2199 was modelled with kT in the 2.5-4
keV range, and abundances decreasing from $\sim$ 0.7 solar at the
center to 0.3 solar for radii $\geq$ 3 arcmin (Siddiqui et al.  1998);
Abell 4059 was modelled with kT increasing from 2 keV at the center to
5 keV at radii $\geq$ 3 arcmin, and abundances in the 0.3-0.6 solar
range (Hwang \& Sarazin 1998); Coma was considered as isothermal with
kT=8.2 keV and abundances 0.21 solar (Hughes et al. 1993); finally,
the hot ICM parameters of Virgo were directly fit to our PSPC data,
with kT in the 1.2-3 keV range, and metal abundances of 0.4-0.6
solar. Using these models, we were able to derive the expected count
rates in the EUV DS band and in the PSPC 1/4 keV band, taking into
account the effective area of the instruments and the performances at
all detector positions. The points corresponding to these calculated
count ratios are shown in the following figures.

In Abell 1795, the EUV to X-ray intensity ratio
(Fig.~\ref{fig:a1795rapp}) is smaller than 1 in the center (for radii
smaller than 1.5 arcmin), but this is most probably due to the
shape of the ratio of the EUV to X-ray PSFs. It then increases with
radius to reach a roughly constant value of 4. The shape of this curve
is consistent with that given by Bonamente et al. (2001b), while the
value of 4 that we find in the outer regions is intermediate between
that of 6 given by Mittaz et al. (1998) and that of 2 given by
Bonamente et al. (2001b). Error bars are typically smaller than $\pm
0.01$ in logarithmic scale, and are therefore difficult to see on the
graphs. We therefore confirm the strong soft excess found in this
cluster by these two groups of authors, and disagree with Bowyer et
al. (1999) who claimed that no excess EUV emission was seen in this
cluster. Note that the small disagreement between the raw and wavelet
reconstructed X-ray image profiles (Fig. \ref{fig:a1795profil}) cannot
be at the origin of the EUV excess that we observe, since on the
contrary the raw X-ray flux is higher that the reconstructed one, and
therefore using the raw X-ray data would lead to an even larger EUV
excess.

The EUV to X-ray intensity ratio in Abell 2199 is smaller than 1 for
radii below 5 arcmin, but as for Abell 1795 this is most
probably due to the shape of the ratio of the EUV to X-ray PSFs. It 
then increases and reaches a value of about 3.7 between radii of 10
and 16 arcmin. We therefore agree with the soft excess observed in
this cluster from various sets of data by Kaastra et al. (1999), and
here also disagree with Bowyer et al. (1999) who found no excess EUV
emission in this cluster.

The EUV to X-ray intensity ratio in Abell 4059 is smaller than 1 for
radii below 2 arcmin (Fig.~\ref{fig:a4059rapp}), in agreement with
Bergh\"ofer et al. (2000b), but again this is most probably due
to the shape of the ratio of the EUV to X-ray PSFs; for increasing
radii, it strongly increases and reaches a value of almost 20 between
radii of 12 and 16 arcmin, in total disagreement with Bergh\"ofer et
al. (2000b), who find no EUV emission beyond 4 arcmin and no EUV
excess beyond 2 arcmin.

Contrary to what happens in the three previously studied clusters, the
EUV to X-ray intensity ratio in Coma appears rather flat up to a
radius of at least 10 arcmin (Fig.~\ref{fig:comarapp}), but as shown
above the two distributions are not similar. The fact that no dip
is seen within the central few arcminutes, as would be expected from
the shape of the ratio of the EUV to X-ray PSFs could indicate that
the EUV excess in the center is even stronger than seen in
Fig. \ref{fig:comarapp}.  There is an EUV excess by a factor of $\sim
1.4$ over a thermal bresmsstrahlung tail.

The EUV to X-ray emission ratio in Virgo is in agreement with the EUV
excesses previously reported by Bergh\"ofer et al. (2000a) and
Bonamente et al. (2001b), therefore implying that here also the EUV
emission is not a thermal bremsstrahlung tail. The rise and fall
of the EUV to X-ray ratio for radii increasing between 0.3 and 2.5
arcmin (Fig. \ref{fig:virgorapp}) is rather surprising since it varies
in an opposite way to the ratio of the EUV to X-ray PSFs; however
images with higher spatial resolution are obviously necessary to push
the analysis further at these radii.

\begin{figure}
\centering
\mbox{\psfig{figure=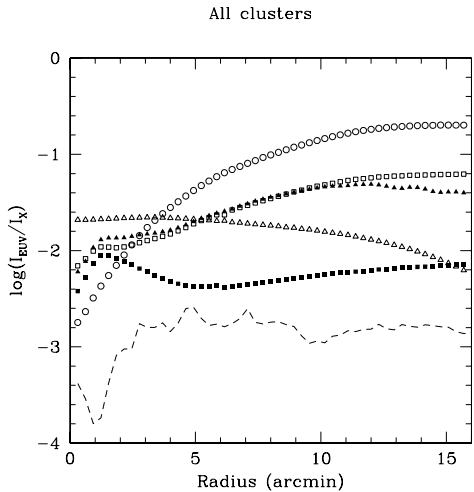,width=7.0cm}}
\caption[]{EUV and X-ray profiles for the five clusters in our sample,
with the radius expressed in arcminutes.  The symbols are the
following: Abell 1795: filled triangles, Abell 2199: empty squares,
Abell 4059: empty circles, Coma: empty triangles, Virgo: filled
squares. Error bars are omitted for clarity. The dashed line
shows the ratio of the EUV to X-ray PSFs.}  
\label{fig:allrminprofil}
\end{figure}

\begin{figure}
\centering
\mbox{\psfig{figure=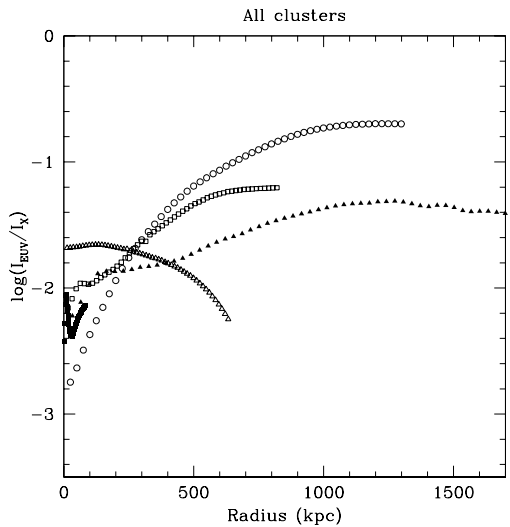,width=7.0cm}}
\caption[]{EUV and X-ray profiles for the five clusters in our sample,
with the radius expressed in kpc.   }
\label{fig:allrkpcprofil}
\end{figure}

The superimposed EUV to X-ray intensity ratios for all the clusters of
our sample are plotted in Figs. \ref{fig:allrminprofil} and
\ref{fig:allrkpcprofil} for radii expressed in arcminutes and in kpc
respectively. Although the shapes of these curves in the central
2.5 arcmin most probably reflect that of the ratio of the EUV to X-ray
PSFs (Fig. \ref{fig:allrminprofil}), we can see that at least beyond
this radius all the clusters do not have the same behaviour. This is
particularly true when radii are expressed in kpc, and certainly must
be linked to a difference in the emission mechanisms of the EUV
emission in these clusters. Note that the radial extent of the EUV
and X-ray emission in Virgo is comparable to the other clusters when
the radius is expressed in arcminutes, but becomes extremely small in
physical distance units, due to the very small redshift of the
cluster: slightly more than 80 kpc.

\section{Discussion and conclusions}

In this paper, we have shown unambigously the existence of a EUV
excess in all five clusters of our sample.

In the first three clusters (Abell 1795, Abell 2199 and Abell 4059),
the EUV to X-ray intensity ratios have comparable behaviours: they
show a possible deficiency of EUV emission over a bremsstrahlung tail
in the very central regions (but this may be an effect of the ratio of
the EUV to X-ray PSFs), and a EUV excess beyond a few arcmin. As
suggested by several authors (Bowyer et al. 1999; Lieu et al. 2000),
the first of these features, if real, can be interpreted as due to
excess absorption within the cluster core due to the fact that in the
cooler central regions some metals are not fully ionized and these
ions absorb part of the soft X-ray flux. The presence of such excess
absorption would agree with the previously claimed existence of
cooling flows in these three clusters. However, XMM-Newton has
detected much weaker emission lines than expected from {\sl bona fide}
isobaric cooling flow models, implying that there is significantly
less ``cool'' gas than predicted (e.g. Kaastra et al. 2001). Therefore
it is no longer straightforward to interpret excess absorption at the
center of these clusters as due to the presence of cooler gas.

One may however note that the only cluster (Coma) for which no EUV dip
is seen in the very center is both the hottest one by far and the only
one with no ``cooling flow'' whatsoever. Since the ratio of the
EUV to X-ray PSFs shows a dip for radii smaller than about 2.5 arcmin,
the absence of such a dip in Coma suggests that in fact there is a
significant EUV excess in the central regions of Coma.

Virgo shows a different behaviour, since the EUV excess is the
strongest in the very center, within radii smaller than about 4 arcmin
(corresponding to a physical size of $\sim$20-30 kpc, that is roughly
the size of M87), then remains roughly constant with radius between 4
and 7 arcmin, and finally increases with a shallow slope for radii
larger than 7-8 arcmin. As for Coma, the presence of a bump in
the central regions suggests that the EUV excess is in fact very
stronge in this zone.  For these last two clusters, images with higher
spatial resolution are obviously necessary to analyze the EUV excess
at these small radii.

The mere presence of an EUV excess in all five clusters of our sample
indicates that a mechanism other than bremsstrahlung is needed to
account for this emission.  Our data does not allow us to discriminate
between the various mechanisms proposed in the literature for the soft
excess. However, in view of the most recent results obtained with
XMM-Newton suggesting that there is much less warm gas in the central
regions of clusters than previously believed, it seems likely that
this EUV excess is probably of non thermal origin.

\begin{acknowledgements}

We are very grateful to M. Traina, from the Observatoire de la C\^ote
d'Azur, for her invaluable help in obtaining the up-to-date version of
the multiscale vision model package used throughout the technical part
of this paper. We acknowledge discussions with D. Gerbal, G. Lima
Neto, G.~Mamon and J. Mittaz. We also warmly thank the referee, Jelle
Kaastra, for several pertinent suggestions which helped us to improve
the paper. Finally, R.~Lieu thanks the Institut d'Astrophysique de
Paris, CNRS, and the Universit\'e Pierre et Marie Curie for their
hospitality.

\end{acknowledgements}


\begin{thebibliography}{}

\bibitem[Anscombe (1948)]{Anscombe} Anscombe, F. J. 1948, Biometrika
15, 246

\bibitem[Atoyan \& V\"olk (2000)]{Atoyan}
Atoyan A.M. \& V\"olk H.J. 2000, ApJ 535, 45

\bibitem[Bergh\"ofer et al. (2000a)]{Berghofer1} 
Bergh\"ofer T.W., Bowyer S. \& Korpela E. 2000a, ApJ 535, 615

\bibitem[Bergh\"ofer et al. (2000b)]{Berghofer2} 
Bergh\"ofer T.W., Bowyer S. \& Korpela E. 2000b, ApJ 545, 695

\bibitem[Bonamente et al. (2001a)]{Bonamente1}
Bonamente M., Lieu R. \& Mittaz J.P.D. 2001a, ApJ 546, 805

\bibitem[Bonamente et al. (2001b)]{Bonamente2}
Bonamente M., Lieu R. \& Mittaz J.P.D. 2001b, ApJ 547, L7

\bibitem[Bowyer \& Bergh\"ofer (1998)]{Bowyer1}
Bowyer S. \& Bergh\"ofer T.W. 1998, ApJ 506, 502

\bibitem[Bowyer et al. (1998)]{Bowyer}
Bowyer S., Lieu R. \& Mittaz J. 1998, in The Hot Universe, ed.
K. Koyama, S. Kitamoto \& M. Itoh (Dordrecht: Kluwer), 185

\bibitem[Bowyer et al. (1999)]{Bowyer3}
Bowyer S., Bergh\"ofer T.W. \& Korpela E.J. 1999, ApJ 526, 592

\bibitem[Bowyer et al. (2001)]{Bowyer4}
Bowyer S., Korpela E.J. \& Bergh\"ofer T.W. 2001, ApJ 548, 135

\bibitem[Briel \& Henry (1996)]{Briel}
Briel U.G. \& Henry J.P. 1996, ApJ, 472, 131

\bibitem[Brunetti et al. (2001)]{Brunetti}
Brunetti G., Setti G., Feretti L. \& Giovannini G. 2001, 
MNRAS 320, 365

\bibitem[Dixon et al. (2001)]{VDD} Dixon W.V., Sallmen
S., Hurwitz M. \& Lieu R. 2001, ApJ 550, L25

\bibitem[Ensslin \& Biermann (1998)]{Ensslin98}
Ensslin T.A. \& Biermann P.L. 1998, A\&A 330, 90

\bibitem[Ensslin et al. (1999)]{Ensslin}
Ensslin T.A., Lieu R. \& Biermann P.L. 1999, A\&A 344, 409

\bibitem[Fabian (1997)]{Fabian}
Fabian A.C. 1997, Science 275, 48

\bibitem[Hughes et al. (1993)]{hughes}
Hughes J.P., Butcher J.A., Stewart G.C. \& Tanaka Y. 1993, ApJ, 404, 611

\bibitem[Hwang (1997)]{Hwang}
Hwang Z. 1997, Science 278, 1917

\bibitem[Hwang \& Sarazin (1998)]{Hwang2}
Hwang Z. \& Sarazin C.L. 1998, ApJ, 496, 728

\bibitem[Kaastra  (1992)]{Kaastra0}
Kaastra J.S. 1992, An X-ray spectral code for optically thin plasmas,
Internal SRON Leiden report v.2.0

\bibitem[Kaastra et al. (1999)]{Kaastra}
Kaastra J.S., Lieu R., Mittaz J.P.D. et al. 1999, ApJ 519, L119

\bibitem[Kaastra et al. (2001)]{Kaastra2}
Kaastra J.S., Tamura T., Peterson J., Bleeker J. \& Ferrigno C. 2001,
Proc. Symposium ``New visions of the X-ray Universe in the XMM-Newton and
Chandra Era'', 26-30 November 2001, ESTEC, The Netherlands

\bibitem[Lieu et al. (1996a)]{Lieu1}
Lieu R., Mittaz J.P.D., Bowyer S. et al. 1996a, Science 274, 1335

\bibitem[Lieu et al. (1996b)]{Lieu2}
Lieu R., Mittaz J.P.D., Bowyer S. et al. 1996b, ApJ 458, L5

\bibitem[Lieu et al. (1999a)]{Lieu3}
Lieu R., Ip W.-H., Axford W.I. \& Bonamente M. 1999a, ApJ 510, L25

\bibitem[Lieu et al. (1999b)]{Lieu5}
Lieu R., Bonamente M. \& Mittaz J.P.D. 1999b, ApJL 517, L91

\bibitem[Lieu et al. (1999c)]{Lieu6}
Lieu R., Bonamente M., Mittaz J.P.D. et al. 1999c, ApJL 527, L77

\bibitem[Lieu et al. (2000)]{Lieu4}
Lieu R., Bonamente M. \& Mittaz J.P.D. 2000, A\&A 364, 497

\bibitem[Mewe et al. (1985)]{Mewe1}
Mewe R., Gronenschild E.H.B.M. \& van den Oord G.H.J. 1985, A\&A, 62, 197

\bibitem[Mallat (1989)]{Mallat}
Mallat S. 1989, IEEE on PAMI 11(7):574-693

\bibitem[Mewe et al. (1986)]{Mewe2}
Mewe R., Lemen J.R. \& van den Oord G.H.J. 1986, A\&A, 65, 511

\bibitem[Mittaz et al. (1998)]{Mittaz}
Mittaz J.P.D., Lieu R. \& Lockman F.J. 1998, ApJ 498, L17

\bibitem[Morrison \& McCammon (1983)]{Morrison}
Morrison R., McCammon D. 1983, ApJ, 270, 119

\bibitem[Ru\'e \& Bijaoui (1997)]{Rue}
Ru\'e F., Bijaoui A. 1997, Experimental Astronomy 7,129

\bibitem[Sarazin \& Lieu (1998)]{Sarazin}
Sarazin C.L., Lieu R. 1998, ApJ 494, L177

\bibitem[Siddiqui et al. (1998)]{Siddiqui}
Siddiqui H., Stewart G.C. \& Johnstone R.M. 1998, A\&A, 334, 71

\bibitem[Snowden et al. (1994)]{Snowden}
Snowden S. L., McCammon D., Burrows D. N. \& Mendehall J. A. 1994, ApJ 424, 714

\end{thebibliography}
\end{document}